\documentclass[11pt]{article}

\usepackage{graphicx}
\usepackage{bbm}
\usepackage{amsmath}
\usepackage{float}
\usepackage{xcolor}
\usepackage{subcaption}
\usepackage[margin=1in]{geometry}
\usepackage[citestyle=authoryear,natbib=true]{biblatex}
\usepackage{setspace}
\usepackage{hyperref}
\usepackage{tabularx}
\usepackage[flushleft]{threeparttable}
\usepackage{soul}
\newcommand{\PreserveBackslash}[1]{\let\temp=\\#1\let\\=\temp}
\newcolumntype{C}[1]{>{\PreserveBackslash\centering}p{#1}}
\newcolumntype{R}[1]{>{\PreserveBackslash\raggedleft}p{#1}}
\newcolumntype{L}[1]{>{\PreserveBackslash\raggedright}p{#1}}

\newenvironment{myindent}
{\par\leftskip1cm\relax\rightskip1cm\relax}
{\par\leftskip0cm\relax\rightskip0cm\relax}

\title{Geographic Spillover Effects of Prescription Drug Monitoring Programs (PDMPs)}
\author{Daniel Guth\thanks{ Guth: California Institute of Technology. Corresponding author, email dguth@caltech.edu. Zhang: California Institute of Technology.}
\and Shiyu Zhang}
\date{\today}

\begin{document}

\begin{titlepage}
\maketitle

\begin{abstract}

    \noindent Prescription Drug Monitoring Programs (PDMPs) seek to potentially reduce opioid misuse by restricting the sale of opioids in a state. We examine discontinuities along state borders, where one side may have a PDMP and the other side may not. We find that electronic PDMP implementation, whereby doctors and pharmacists can observe a patient's opioid purchase history, reduces a state's opioid sales but increases opioid sales in neighboring counties on the other side of the state border. We also find systematic differences in opioid sales and mortality between border counties and interior counties. These differences decrease when neighboring states both have ePDMPs, which is consistent with the hypothesis that individuals cross state lines to purchase opioids. Our work highlights the importance of understanding the opioid market as connected across counties or states, as we show that states are affected by the opioid policies of their neighbors.     

    %\vspace{0in} \\ 
    %\noindent\textbf{Keywords:} Opioids, Drug Overdoses, Drug Distributors\\
    %\vspace{0in}\\
    %\noindent\textbf{JEL Codes:} I11, I12, I18 \\
    %\bigskip
\end{abstract}
\end{titlepage}

% \section*{Changes from last time}
% \begin{enumerate}
%     \item All the edits from JL are incorporated. Changes that might need additional review are highlighted.
%     \item Shiyu did significant editing to the hypothesis section: (1) clearly stated the testable hypotheses, (2) added 2 paragraphs on when are individuals incentivized to cross the border and how ePDMP changes that, and (3) added a figure to summarize the hypotheses under the spillover framework
%     \item Section 5.3 on alternative dates is new
    
% \end{enumerate}

% \section*{Remaining Issues}
% \begin{enumerate}
%     \item One final round of proof-reading
%     \item I can't think of a good way to empirically show that endogeneity of adoption bias the coefficients toward no effect (attempted in Table\ref{tab:endogeneity}). So i don't think this is a statement we could make. Without this statement, paragraph 2 on page 7 needs more work.
%     \item A couple of TODO sprinkled around the paper
% \end{enumerate}

\newpage

\section{Introduction}
Over the past two decades, the opioid epidemic has claimed more than 415,000 American lives (CDC Wonder). To stem the rising tide of opioid misuse, in the early 2000s, states began to regulate prescription opioid sales. Among the different policies that were implemented, we focus on Prescription Drug Monitoring Programs (PDMPs) that require prescribers and dispensers to submit data to a centralized system. In this paper, we study the effects of states' implementation of electronic-access PDMPs, a version of the law that allows doctors and pharmacists to query the patient's prescription history in real-time, on different regions in the same state and on the nearby states. Specifically, we focus on how sales in counties that border other states react differently to new PDMP regulations from sales in `inland' counties.

Our analysis shows that electronic-access PDMPs reduce prescription opioid sales and opioid mortality. The effect is economically and statistically significant despite the fact that endogenous adoptions of such regulations bias our estimates of their impact downward. We find that border counties (counties that are immediately adjacent to another state) are systematically different from inland counties (counties not immediately adjacent to a county in a different state) and the enactment of ePDMP laws disproportionately affects border counties. These findings are consistent with our hypothesis that the border counties are destinations for consumers who are doctor or pharmacy shopping due to their proximity to another state. We also find a small but significant spillover effect in the form of increased opioid sales and overdose deaths when the neighboring state adopts stricter PDMP regulations.

Using the novel ARCOS data, we confirm the literature's general finding that PDMPs reduce opioid sales and mortality. We also contribute to resolving a debate in the literature about what features of PDMPs are more effective than others. We find that one specific implementation, electronic-access PDMPs (ePDMPs), are most effective at reducing opioid sales and mortality. Compared to a regular PDMP, this version not only requires doctors to submit information, but also allows doctors to see what other opioids a patient has received in real time. To the extent doctors and pharmacists consult the databases, ePDMPs mitigate the problem of individuals going to multiple doctors or pharmacies to secure opiates. We find that ePDMP laws reduce per person sales by 0.006 mg in activate ingredient weight\footnote{The active ingredient weight is equivalent to the morphine milligram equivalent (MME) divided by 1500}, which is equivalent to a 5.6\% drop the 2006 national average, and per 100,000 mortality by 0.279, which is equivalent to a 12.3\% from the 2006 national average.

We perform our analyses at the county level, which allows us to measure systemic differences in opioid markets of inland versus border counties due to the presence of state borders. Border counties appear similar to inland counties on observable demographics, but they have significantly higher opioid sales and lower opioid overdose deaths. These differences are consistent with the hypothesis that border counties are more frequently the destinations of doctor or pharmacy shopping, largely because their proximity to other states leads to lower travel costs for out-of-state residents. This difference between border counties and inland counties falls after the state adopts ePDMP, which further confirms our hypothesis that a higher percentage of sales in border counties were trafficked elsewhere for consumption. Our findings challenge the states-as-islands model often assumed in the opioid literature. 

%We urge the literature take into consideration these differences and potential spatial spillover in evaluation of policy effectiveness.

We also document negative externalities from these ePDMP laws, in the form of opioid sales and mortality increases in the border counties of neighboring states. We argue that these externalities come from the demand-side response of individuals using opioids, who now acquire these prescription drugs from out-of-state. The substitution to opioids from other states potentially reduces the effectiveness of ePDMPs as a policy intervention. The spatial substitution identified in this paper builds upon our previous work \parencite{zhang2021oxycontin} showing that partial supply-side interventions, like the OxyContin reformulation, can lead to drug substitution instead of preventing misuse. In the case of ePDMPs, the policy intervention was at the state and not the national level, so sales shifted across state lines instead of across products. This paper adds to the growing literature on the side effects of supply-side intervention curbing the opioid crisis (\citealp{alpert2018supply}; \citealp{kim2021must}).

Our work speaks to the importance of not analyzing individual state policies in a vacuum. Individuals frequently cross these invisible borders in their day-to-day lives, and they may thus be subject to different regulatory regimes. The ability for individuals to evade one state's regulations for another extends to all markets regulated at the state level. One of the policy implications of our work is that there are costs to regulating opioids at the state level, and there would be benefits in enacting a national ePDMP. The American College of Physicians has called for a national prescription drug monitoring program, and for standardized PDMP laws across states until that point \parencite{kirschner2014prescription}.

The rest of the paper is as follows. Section 2 gives a background on PDMP laws as well as an overview of the literature understanding their effects. Section 3 describes the county-level sales and mortality data we use, the spread of PDMP laws during this time period, and our categorization of border counties. Section 4 describes how we model PDMP-border counties as well as our predictions based on economic theory and known trafficking patterns. Section 5 provides our results on sales and mortality, and finally, Section 6 concludes.

\section{Background and Literature Review}

Our paper connects three different strands of literature. First, we contribute to the literature on the opioid crisis and policies curbing opioid misuse. Second, we take methods from spatial economics and apply them to cross-border opioid sales and misuse. Third but not least, we build upon modern analyses of the effects of PDMP laws. 

\subsection{The Opioid Crisis and Interventions}

Over the past two decades, millions of Americans have misused prescription opioids. In 2019 alone, 1.6 million people had an opioid use disorder and 70,630 people died from an opioid overdose\footnote{The US Department of Health and Human Services on the Opioid Epidemic \href{https://www.hhs.gov/opioids/about-the-epidemic/index}{(link)}}. Opioid use disorder has devastating consequences for the individual, the family, and the community. The CDC estimates the total ``economic burden'' of prescription opioid misuse to be 78 billion dollars a year. 

Many victims of the epidemic got their first access to an opioid from a doctor's prescription. Previous research has documented large variations in opioid prescribing and sales, both within and across states. \citet{mcdonald2012geographic} shows that the ratio of per-capita oxycodone sales in counties in the 75th percentile to counties in the 25th percentile is approximately 7 to 1. Their best model can only predict one-third of the variation in sales by county. \citet{finkelstein2018drives} uses Medicare data to track individuals who move between counties, and the paper finds that location has a noticeable effect on an individual's access to opioids. The paper estimates that 30\% of the difference in opioid prescribing between counties can be explained by these place-specific factors. Our work is connected to the opioid prescription literature, in that we both study location-specific effects, but our data is on opioid shipments to pharmacies which occurs further down the prescription pipeline. We add to this literature by showing that being on the state border is one of these factors that affect local opioid sales and misuse. 

Over the last two decades, states have made repeated attempts to regulate the sales of prescription opioids in the hope of preventing further opioid misuse. Litigation against Purdue Pharma, the manufacturer of the drug that ignited the opioid crisis, led the company to reformulate OxyContin in 2010. The reformulation led to reduced sales of OxyContin, but spurred on an increase in alternative oxycodone and heroin misuse (\citealp{zhang2021oxycontin}; \citealp{alpert2018supply}; \citealp{evans2019reformulation}). Many states enacted new PDMP laws or tightened existing ones. The evidence of the effectiveness of such laws is mixed (for more detail see Section \ref{section:bg}), and some argue that the new restrictions led to increases in heroin mortality (\citealp{kim2021must}; \citealp{Dave2021}). Some states, Florida included, passed legislation that requires pill mills---rogue pain management clinics that were inappropriately prescribing and dispensing opioids---to register with the state. The pill mill laws have led to a moderate decrease in opioid prescription and use (\citealp{rutkow2015effect}; \citealp{kennedy2016opioid}). One common theme in this strand of the literature is a substitution toward alternative drugs when the original supply became restricted. We add to this literature by evaluating the effectiveness of PDMP laws while taking into consideration potential spatial spillovers. 

\subsection{Spatial Spillover and Opioids}

Our work ties tightly into the literature studying the distribution of economic activities across space. Many works have noted how geographic characteristics have a direct impact on manufacturing, sales, and trade. \citet{holmes1998effect} finds sharp increases in manufacturing activity across the border in so-called 'pro business' states. Similarly, \citet{nachum2000economic} finds that location and agglomeration effects can explain which states transnational corporations choose to put their headquarters in. \citet{fox1986tax} examines border counties and finds that changes in state taxes can shift purchases across state lines. \citet{garrett2002revenue} examines lottery sales in Kansas and estimates that the state loses \$10.5 million dollars in net lottery revenue to cross-border shopping in 1998. We use border counties, a concept from this literature, to show how state policy differentially affects different locations. Our setting provides the perfect environment to test for spillovers because we have detailed sales data on exactly where opioids are sold, which is not common in other settings.

We also contribute to a small but significant literature on cross-border prescription shopping. Crossing state and national borders to taken advantage of favorable regulatory environments to obtain drugs is not a new concept in the literature. \citet{casner1992purchasing} documents patients crossing the US-Mexico border to purchase prescription medication cheaply and without a prescription. \citet{mcdonald2014ecology} estimates that approximately 30\% of ``doctor shoppers" had opioid prescriptions from multiple states. \citet{cepeda2013distance} finds that 4\% of non-shoppers visited more than one state to purchase opioids, and for individuals who visited multiple pharmacies to purchase opioids, the median distance between pharmacies was about 12.6 miles. We add to this literature by leveraging decentralized policy change to systematically identify the impact of cross-border shopping on opioid sales.

\subsection{Prescription Drug Monitoring Programs}\label{section:bg}

Before any Prescription Drug Monitoring Programs, or PDMPs, individuals can freely doctor or pharmacy shop\footnote{Doctor shopping refers to the behavior of individuals going to multiple doctors to get opioid prescriptions to evade scrutiny, and pharmacy shopping refers to going to multiple pharmacies to get the prescriptions filled}, and there is no way for doctors or pharmacists to know how many other prescriptions an individual has. PDMPs are state-level databases that track controlled substance prescriptions in a state. The modern precursor to the PDMP was California's `Triplicate Prescription Program' enacted in 1939\footnote{New York had the first PDMP law in 1918, but rescinded it three years later.}. The law required the dispensing pharmacist to fill out standardized forms for controlled substances and mail a copy to a centralized state repository. The California program set the blueprint for PDMPs and many states followed suit in subsequent decades. The legality of PDMPs was tested in \textit{Whalen v. Roe}, where the Supreme Court unanimously ruled that storing this personal medical information did not violate a person's right to privacy. 

These original PDMPs collected information from doctors and pharmacists via mail or fax, and doctors and pharmacists could not immediately query a patient's opioid history. Oklahoma implemented the first fully electronic PDMP in 1990 that directly and routinely sent records to a state database \parencite{holmgren2020history}. Currently, the electronic-access PDMPs allow registered doctors and pharmacists to query the data set in real-time and see all opioids an individual received in that state.  The 21st century saw a wave of expansion to electronic PDMPs and by 2019, all but one US state have implemented e-access PDMPs \parencite{mallatt2019unintended}. The next wave of PDMP regulation is the must-access or mandatory PDMPs. These laws require doctors and pharmacists to check an individual's opioid history before dispensing opioids. Absent the mandate, only filling the information is mandatory; checking a patient's history is voluntary. The must-access laws are often based on, and enacted after, electronic PDMPs. By 2017, 19 states have enacted some version of must-access PDMPs. 

Most states do not share any information collected from PDMPs with other states\footnote{\cite{lin2019interstate} shows that in 2014, 23 states had some sort of data sharing agreement, but many of these agreements were one-way, and only Michigan and Indiana shared this information with all of their neighboring states.}. The lack of information sharing made it feasible for individuals to partially circumvent the regulation by shopping across state borders. If state A adopted an ePDMP, an individual would face greater difficulty getting a second opioid prescription filled in-state. This difficulty could occur either from doctors, who upon observing that a patient already has an opioid prescription do not write another, or from pharmacists who refuse to fill it for the same reason. However, an individual could attempt to get and fill a second prescription in a neighboring state. We aim to evaluate the propensity for individuals to get opioid prescriptions outside of their state, specifically to avoid PDMP regulations.

There is a wide array of studies on the effects of PDMPs. One typical corroborated result in the literature is that PDMPs decrease prescription opioid sales (\citealp{Simeone2006}; \citealp{Reisman2009}; \citealp{Kilby2016}) and reduce abuse and mortality (\citealp{Simeone2006}; \citealp{Patrick2016}). Some papers note that specific formulations of PDMP are more effective than others. Effective features include monitoring more drugs and updating weekly \citep{Patrick2016}, and identifying and investigating cases proactively \citep{Simeone2006}. \citet{bao2016prescription} look at 22 states from 2001 to 2010 that implemented electronic access to PDMPs and showed it reduced oxycodone prescriptions from ambulatory visits to physician offices by 30\%. A set of papers claim that only must-access PDMPs (MA-PDMPs) are effective in reducing opioid misuse (\citealp{Buchmueller2018}; \citealp{grecu2019mandatory}; \citealp{Dave2021}; \citealp{Meinhofer2018}; \citealp{kim2021must}) which conflicts with existing results on effectiveness of non-mandatory PDMPs. We contribute to this debate by showing that ePDMP laws are effective at reducing sales and overdose deaths during our sample period. 

The disagreement in the literature on what features of PDMPs are more effective than others is partly the result of each paper employing its own categorization of laws and testing the effectiveness on different outcome variables. Assembling an accurate policy data set across all 50 states is inherently challenging \citep{schuler2021methodological}. \citet{horwitz2018problem} point out that the inconclusive and contradictory results may be due to the large variations in dates used in different studies. Existing sources of enactment dates rarely acknowledge the researchers' decisions in creating such a data set, and the public sources have a large disagreement. In this paper, we use the ``modern system operational date'' variable from \citet{horwitz2018problem} in our main analysis. We will elaborate on the choice of ``modern system operational date'' over other implementation dates in Section \ref{sec: enactment dates}. 

States adopted PDMP policies at different times, but the literature generally does not address potential endogeneity concerns. For our regression specification, one particularly worries that states might be more likely to adopt PDMPs because they have the infrastructure to make the laws effective. If so, a naive regression's coefficients would be biased in favor of the hypothesis that the laws matter but such upward biases are unlikely. We argue that adoptions of PDMP laws are endogenous to local conditions but in ways that bias coefficients downwards towards zero rather than upwards. Specifically, places that are experiencing more opioid misuse or higher growths in sales or overdoses are the most likely to adopt measures like PDMPs. A simple difference-in-difference estimation of the effect of the law underestimates its impact and biases against the key hypotheses we want to test. Our estimation of the impact of PDMP laws on sales and mortality both suffer from this bias, but we are capturing statistically significant coefficients nonetheless. Moreover, since the enactment of PDMPs in a state is independent of the differences between border and inland counties in that state, and independent of conditions in nearby states, our estimation of the border effect and spillover will not be affected by the endogeneity problem.

% and examine changes in opioid sales and mortality. \citet{kaestner2019mortality} has previously shown that a version of that variable reduces hydrocodone and oxycodone sales in a state. 

%These papers focus on substitution to illegal channels, but we show substitution to out-of-state markets.  \textbf{We add to this literature by showing that ePDMP policies are effective at reducing sales and overdose deaths in a state, and thus that there is some benefit to these laws even without mandatory access.}

%Several papers (\cite{kim2021must}, \cite{Dave2021}) document a rise in heroin deaths in response to states passing must-access PDMPs. This proposed channel is that by cutting off the supply of prescription opioids, it induces individuals to switch to heroin. We build on this by identifying another spillover channel, purchasing opioids across state lines. \textbf{We test several different implementation dates, including the MA-PDMP dates from \cite{mallatt2018effect} used in \cite{kim2021must} as robustness checks.}  \textcolor{blue}{Not entirely sure where to put this paragraph} 

\section{Data}\label{section:data}

In this section, we introduce the data source of our sales and mortality data, describe our choice of PDMP implementation dates, define how we characterize border counties, and present summary statistics.

\subsection{ARCOS sales data and NVSS Mortality data}

As part of the Controlled Substances Act, distributors and manufacturers of controlled substances are required to report all transactions to the DEA. This Automation of Reports and Consolidated Orders System (ARCOS) database contains the record of every pain pill sold in the United States. The complete database from 2006 to 2014 was recently released by a federal judge as a result of an ongoing trial in Ohio against opioid manufacturers.\footnote{\href{https://www.washingtonpost.com/graphics/2019/investigations/dea-pain-pill-database/}{Link} to the ARCOS Data published by the Washington Post.} The part of ARCOS that we use in this paper is shipments of oxycodone from manufacturers to pharmacies. In theory, the manufacturer to pharmacy shipments are not equivalent to sales to the consumers. However, since pharmacies do not keep large stocks of opioids, the aggregated annual data of sales from manufacturers to pharmacies is practically equivalent to the annual sales of pharmacies to consumer sales. The benefit of ARCOS data is that it allows disaggregation to arbitrarily fine geographical levels, which is essential for the identification of the border effects, and it contains all opioid sales which allows us to identify spatial substitution. The ARCOS sales data is the primary outcome variable in our regressions.

We care about how PDMP laws affect opioid sales, but ultimately we're interested in preventing their effects on overdoses and deaths. The second outcome of interest in our main regression is opioid mortality. We use the restricted-use multiple-cause mortality data from the National Vital Statistics System (NVSS) to track opioid overdoses. The dataset covers all deaths in the United States from 2006-2014. We follow the literature's two-step procedure to identify opioid-related deaths. First, we code deaths with ICD-10 external cause of injury codes: X40–X44 (accidental poisoning), X60–64 (intentional self-poisoning), X85 (assault by drugs), and Y10–Y14 (poisoning) as overdose deaths. Second, we use the drug identification codes, which provide information about the substances found in the body at death, to restrict non-synthetic opioid fatalities to those with ICD-10 code T40.2.

\subsection{PDMP Enactment Dates} \label{sec: enactment dates}

As discussed in the background section, there are multiple sets of PDMP enactment dates and the literature disagrees on which is the most effective in reducing opioid misuse. In this paper, we consider three sets of dates: (a) the legislated start date (any PDMP), which is the year that dispensers or prescribers would be required to send prescriptions to a central database, (b) the electronic access date (ePDMP), which is the year that the PDMP data becomes accessible to the dispensers or prescribers through a centralized electronic system, and (c) the must access date (MA-PDMP), which is the year when certain dispensers or prescribers are required to check an individual's opioid history before dispensing. In Figure \ref{fig:implementation} in Appendix, we graph the three enactment dates for each state. Most states started with the most basic version of PDMP and gradually adopted e-access in the 2000s. Only a handful of states adopted must-access PDMP during our time period.

We use ePDMP dates in our main regression analysis. The reasons are twofold. First, ePDMPs have large impact on prescriptions and sales both conceptually and empirically. Conceptually, an ePDMP streamlines the process by which the prescribers and dispensers check a patient's prescription history. Before an ePDMP, prescribers and dispensers are required to report opioid prescriptions but could not easily tell what other prescriptions an individual had. ePDMPs allows them to check a patient's opioid history online in real-time, so they could more easily refuse opioids to questionable patients. Although an ePDMP is less restrictive than a MA-PDMP, it is reasonable to assume a large number of doctors who are conscious of the severity of the opioid crisis would have taken advantage of the electronic system when it became available. Empirically, \cite{horwitz2018problem} finds this set of dates is most correlated with reductions in opioid sales after comparing it with 9 other sets of dates\footnote{To be fair, the paper did not compare ePDMP with MA-PDMP}. 

Second, our sample period has higher coverage of enactment of ePDMP as compared to the other two dates. There is reasonable consensus in the literature that each wave of new PDMP legislation tightens the legal supply of opioids and reduces misuse (although the literature disagrees on which version is the most effective). Given that each round of legislation may have some impact, we want to work with the one that gives us the most identification power. The switch from no PDMP to any PDMP happened in the 1990s and early 2000s, and by 2006, the start of our sample period, 31 states have already adopted it. The adoption of ePDMPs took place mainly during our sample period: 37 states adopted ePDMP between 2006 and 2014. Only 10 states enacted MA-PDMP during our sample period. Working with ePDMPs allows us to use data from more states to estimate the impact of the law. Our $\beta$ estimations would be less reliant on trends from a few states.

Our ePDMP dates are obtained from \citet{horwitz2018problem}\footnote{The authors coined their e-access dates the ``modern system operational date''. Although the naming is different, the two definitions are conceptually identical.}. We Horwitz as our main source because this paper is the most systematic methodological paper on PDMP implementation timing that we've reviewed. In robustness, we use ePDMP dates published by the Prescription Drug Abuse Policy System (PDAPS), an organization funded by the National Institute on Drug Abuse to track state laws related to prescription opioid abuse. To check if other PDMP laws have similar spillover effects, we use any PDMP dates from \citet{horwitz2018problem} and MA-PDMP dates from \citet{sacks2021can}. We list all sets of dates in Table \ref{tab:date} in the Appendix. 

\subsection{Defining Border Counties and Assigning ePDMP Status}

We define a border county as a county that neighbors at least one county in an adjacent state and an inland county as a county that borders only counties of the same state. After excluding Alaska, Hawaii\footnote{Alaska and Hawaii neighbor no US states.}, and Florida\footnote{Florida experienced a dramatic rise in opioid supply in the 2000s and then a significant drop due to crackdown on pill mills in 2010–2011. It is common practice in the literature to exclude Florida from the analysis.} from our data, we have 2906 counties, 37.3\% of which are border counties (see Figure \ref{fig:map_border} in Appendix for a visual representation of border and inland counties). For each inland county, we document whether an ePDMP law has been implemented. For each border county, we document whether a law has been implemented in that state and the bordering state(s). If a county is bordering multiple states and these states have different ePDMP status, the nearby law of the county will be the ePDMP status under which the majority of the nearby population live\footnote{The underlying idea is that the ePDMP status of more populous nearby counties would have a bigger impact on my county than the ePDMP status of less populous nearby counties. Specifically, we sum up the population adjacent to a border county by ePDMP status. If more nearby population resides under the states with PDMP law than no law, the county's nearby law variable will be 1; if more nearby population resides under the states with no law, it will be 0.}. We only need to do this calculation on 653 county-year observations, which is 6.6\% of all border county-year observations. See Figure \ref{fig:nearby_example} in Appendix for an example of the calculation. 

The transition from states not having an ePDMP to having an ePDMP is key to our identification. During our sample period, over 60\% of all counties transitioned from no ePDMP to ePDMP (see adoption rate in Figure \ref{fig:pdmp_overtime} in Appendix). Identification of border coefficients relies on law change in a county and law changes in nearby border counties. The majority of the transitions in border counties also took place during our sample period: over 80\% of border counties has no ePDMP regulation in 2006 and that number decreases to less than 20\% by the end of 2014 (see detailed transitions in Figure  \ref{fig:pdmp_border_overtime} in Appendix).

\subsection{Summary Statistics}

Since we are comparing border counties to inland counties of the same state, it is important that we acknowledge any potential differences between the two sets of counties, especially those associated with opioid use. In Table \ref{tab:summary_stat}, we document the population-weighted average of opioid sales, mortality, and important demographics, and ePDMP coverage of the two sets of counties. Border counties have a significantly higher level of opioids sales throughout the sample period. They have lower levels of opioid mortality in 2006, but the difference loses significance since 2010. We will discuss these differences in outcome variables in our hypotheses and result section. The two sets of counties are quite similar on all demographic dimensions. Since some of these demographic factors are associated with higher levels of opioid misuse, it is important we control for demographic differences in our regressions. 

\begin{table} [H]
	\centering
	\footnotesize
	\renewcommand\arraystretch{1.2}
	\begin{threeparttable}
		\caption{Summary Statistics}
\label{tab:summary_stat}
	
\begin{tabular} {L{0.001\textwidth}L{0.3\textwidth}C{0.1\textwidth}C{0.1\textwidth}C{0.1\textwidth}C{0.1\textwidth}}
\hline\hline
 \multicolumn{2}{l}{Variables} & All counties & Border counties & Inland counties & Test of equality (p-value) \\ \hline
 \multicolumn{6}{l}{\textit{Opioid-related statistics}} \\
 & Sales per person (2006) &0.101&0.113&0.095&3.28e-11\\
 & Sales per person (2010) &0.163&0.185&0.150&0.004\\
 & Sales per person (2014) &0.158&0.181&0.145&0\\
 & Opioid overdose per 10,000 (2006) &2.22&2.01& 2.33&0.003 \\
 & Opioid overdose per 10,000 (2010) &3.35&3.29&3.38&0.574 \\
 & Opioid overdose per 10,000 (2014) &3.84&4.02&3.75&0.164 \\
 \multicolumn{6}{l}{\textit{Demographics (2009)}} \\
 & Population &98,853&92,914&102,392&0.397\\
 & Average Age &36.11&36.8&35.7&0.149\\
 & Male (\%) &49.2&49.0&49.3&1.53e-07\\
 & Separated (\%) &18.2&18.6&18.1&0.001\\
 & High School and above (\%) &83.4&83.9&83.1&0.002\\
 & Bachelor and above (\%) &27.4&27.2&27.6&0.004\\
 & Mean income &70,130&71,063&69,625&0.05\\
 & Low income (\%) &33.2&33.3&33.2&0.703\\
 & White (\%) &78.6&79.0&78.4&0.279\\
 & Black (\%) &12.8&13.6&12.4&0.015\\
 & Asian (\%) &4.94&3.87&5.51&0\\
 & Native American (\%) &0.178&0.141&0.197&8.41e-05\\
 \multicolumn{6}{l}{\textit{PDMP-related statistics}}\\
 & Number of counties &2906&1085&1821&\\
 & Have ePDMP by 2006 (\%) & 18.6 & 17.8 & 20.1&\\
 & Have ePDMP by 2010 (\%) & 50.5 & 52.0 & 49.5&\\
 & Have ePDMP by 2014 (\%) & 87.2 & 85.1 & 88.3&\\
 \hline\hline
 
\end{tabular}

\begin{tablenotes}
	\small
	\item \textit{Notes:} Means are weighted by county population. For opioid-related statistics, border counties have significantly higher levels of opioid sales throughout the sample period. Mortality is higher in inland counties, but the difference is not significant in all three years we tested. Many of the differences in demographics between border and inland counties are statistically significant but not economically. The adoption rates of ePDMP laws are similar between the two sets of counties.
\end{tablenotes}
	\end{threeparttable}
\end{table}

%There are three ways one county could transition from no law to law: from $(0,0)$ to $(0,1)$ as the first adopter, from $(1,0)$ to $(1,1)$ as the second adopter, and from $(0,0)$ to $(1,1)$ when adoption occurred simultaneously. Whenever a county transitions from $(0,x)$ to $(1,x)$, some county(s) on the other side of the border would transition from $(x,0)$ to $(x,1)$, which is why the share of $(0,1)$ and $(1,0)$ counties are relatively equal over time. In our main regression, we model how law and nearby law change impacts opioid use patterns in a border county, but not how they interactively do so. In robustness, we discuss if being the first mover or the second mover changes how a border county reacts to law change.

\section{Hypotheses and Empirical Framework}\label{section:model}

\subsection{Hypotheses}

In this section, we lay out our hypotheses and discuss the underlying assumptions and their implications on the market structure of prescription opioids. We start with a simplified model with no spatial spillover.

\textbf{The state-as-island model.} Consider states as isolated islands in an ocean. Due to the separation, opioids sold in each state can only be consumed in that state. Since county location bears no significance in this model, sales patterns and mortality should be similar in border and inland counties of the same state after controlling for demographic differences. For example, San Bernadino County, on the state border between California and Arizona, should behave similarly to Fresno County, landlocked within California. Although the adoption of a PDMP is endogenous to local conditions, a priori we would not expect the law to have differential effects on border and inland counties. Since all opioids sold locally are consumed locally, changes in sales due to PDMP laws should translate directly to changes in use patterns, and by extension, to changes in local opioid mortality, ignoring any substitution to illegal opioids\footnote{We focus on opioid mortality, but as described in the literature review, some papers do find substitution to heroin following implementation of MA-PDMP laws.}. The adoption of PDMP in one state should have no impact on opioid sales or mortality in the neighboring state. The testable hypotheses of the state-as-island model are:

%\hl{[Should we call them hypothesis or implication?]}

\begin{myindent}
\textit{Hypothesis 1a:} Under the state-as-island model, sales and mortality patterns are similar in border and inland counties.
\end{myindent}

\begin{myindent}
\textit{Hypothesis 1b:} Under the state-as-island model, changes in sales translates into changes in mortality.
\end{myindent}

\begin{myindent}
\textit{Hypothesis 1c:} Under the state-as-island model, adoption of PDMP in one state has no impact on sales or mortality in the neighboring states.
\end{myindent}

However unrealistic the above model is, it is assumed in many important studies on the opioid crisis. States are treated as isolated markets where all pills sold are consumed locally with the exception of Florida, which most papers exclude. The state-as-island model is applicable in situations when the spillover effect is small compared to the main effect, or if the spillover's impact is tangential to the main question. The literature has documented many occasions when the state-as-island model fails. Individuals cross the state border to take advantage of favorable lottery situations \parencite{garrett2002revenue}; patients cross the US-Mexico border to purchase prescription medication cheaply and without a prescription \parencite{casner1992purchasing}. The decentralized enactment of PDMP creates differences in regulatory environments and incentives individuals to seek out the less regulated market. Next, we consider a model with spatial spillover.

\textbf{The spatial spillover framework.} Consider two states not separated by an imaginary ocean. Both opioids and people can cross the state line. As a result, opioids purchased in one state may or may not be consumed in that state. When individuals are incentivized to purchase opioids from a neighboring state, their cost of doing so is highly dependent on the distance traveled. Under these assumptions, vicinity to the state border has consequences on opioid sales and diversion. For someone living on the Arizona side of the Arizona-California state border, the cost of travelling to San Bernadino County for additional pills is much lower than that of travelling to some inland county within California. 

%Under this framework, for counties in the same state, how opioids are sold and consumed depends on the county's location within that state. Subsequently, relations between sales and mortality and reactions to PDMP laws will be different for two sets of counties.

The question remains as to when are individuals incentivized to cross the state border? Before any PDMP law, patients could obtain multiple prescriptions and get them filled in the same state with minimal constraint. When states adopt some version of the PDMP, doctor and pharmacy shopping within the same state become more difficult. However, because most states do not share their PDMP data with the neighboring states, the cost of obtaining additional pills from the neighboring states remains the same despite enactment of PDMP locally. As the cost of within-state pill shopping increases due to progressively stricter PDMP regulations (from PDMP to ePDMP to MA-PDMP), more and more individuals would be incentivized to cross a nearby border. By the start of our time period, 31 states had enacted some version of the PDMP, which means that some individuals would already be going to other states for pills. Hence, we expect a higher share of the border counties' sales to be diverted elsewhere for consumption during our sample period. Because the diverted pills are not consumed locally, we expect the sales to mortality ratio to be higher in a border county.

\begin{myindent}
\textit{Hypothesis 2a:} Under the spillover framework, border counties will have higher sales but lower moralities as compared to inland counties of the same state.
\end{myindent}

Variation in diversion rates between inland and border counties implies that the two sets of counties will respond differently to new PDMP regulations. When states enact stricter PDMPs, the local pill shoppers and the out-of-state pill shoppers are similarly affected by the new rule. Since a higher share of the border counties' sales is from pill shoppers, the law change will have a bigger impact on the border counties. The endogeneity of adoption may bias the overall estimation toward zero, but should not affect how the border counties react to the law change relative to the inland counties. In addition, as the cost of local pill shopping increases due to stricter laws, local pill shoppers are more incentivized to cross the state border, and hence sales in border counties of the neighboring states would increase.

\begin{myindent}
\textit{Hypothesis 2b:} Under the spillover framework, when the local state adopts a stricter PDMP, border counties will experience a larger decrease in sales relative to inland counties of the same state.
\end{myindent}

\begin{myindent}
\textit{Hypothesis 2c:} Under the spillover framework, when the nearby state adopts a stricter PDMP, border counties will experience a larger increase in sales relative to inland counties of the same state.
\end{myindent}

In this stylized model, the mapping from sales to mortality is less direct when spatial spillover was not possible. With the state-as-island model, the enactment of a PDMP law puts a hard constraint on the opioid misuser's ability to acquire prescription opioids. Assuming no other substitution, changes in opioid sales in one location translate directly into changes in opioid mortality in that location. With spatial spillovers, changes in opioid sales in one place may lead to changes in mortality elsewhere. Since a larger share of the border counties' sales was consumed elsewhere, the adoption of stricter PDMP will result in a smaller drop in opioid mortality in the border counties. The enactment of PDMP in a nearby state increases sales in the border counties but should have no additional impact on mortality, assuming that people traveling to acquire pills go back to their home counties to consume them. In reality, how mortality responds to a PDMP law depends on many factors, including the state of the black market, the availability of alternative drugs, and the ease of getting drugs from the nearby states. Since we cannot control for all of these relevant factors, we expect the mortality results to be less sharp than the sales results.

\begin{myindent}
\textit{Hypothesis 2d:} Under the spillover framework, when the local state adopts stricter PDMPs, border counties will experience a smaller decrease in mortality relative to inland counties of the same state.
\end{myindent}

\begin{myindent}
\textit{Hypothesis 2e:} Under the spillover framework, when the nearby state adopts stricter PDMPs, border counties will experience no additional change in mortality relative to inland counties of the same state.
\end{myindent}

See Figure \ref{fig:hyp} in Appendix for a visual representation of the hypotheses of the spillover framework.

\subsection{Empirical Framework}

We want to test (1) how counties react to the enactment of ePDMP laws, (2) if border counties react differently as compared to inland counties, and (3) how the adoption of an ePMDP in one state affects border counties in the adjacent state. We use the following empirical framework to test our hypothesis:
\begin{align*}
Y_{ct} = & \alpha_s + \delta_t + \beta_1\text{ Law}_{ct} + \beta_2\text{ Border}_c + \beta_3 \text{ Law}_{ct} \times \text{Border}_c + \\
&\beta_4 \text{ Nearby Law}_{ct} \times \text{Border}_c +  X_{ct}\gamma + \epsilon_{mt}
\end{align*}
where $Y_{ct}$ are the outcome variables of interest: sales and mortality in county $c$ in year $t$. Ideally, because each county has different initial conditions, we want to control for these conditions to get at the impact of the law change. However, because the location of a county and its border status does not change over time, any time-invariant differences between the border and inland counties would be absorbed by the county fixed effects if added. Hence, we use a full set of state fixed effects $\alpha_s$ and county characteristics $X_{ct}$ as controls. We also add year fixed effects to control for national changes in drug use over time. 

%The border coefficients $\beta_2$ and $\beta_3$ capture the within-state differences in reaction to ePDMP laws. These two coefficients measures how border counties reacted differently to ePDMPs as compared to inland counties of the same state. 

Our coefficients of interest are the full set of $\beta$'s: $\beta_1$ estimates the impact of ePDMP laws on sales and mortality; $\beta_2$ estimates the baseline difference in sales and mortality between border and inland counties of the same state; $\beta_3$ estimates how the law affects the border counties differently as compared to the inland counties; and $\beta_4$ estimates how the enactment of an PDMP in one state impacts sales and mortality in the bordering counties of the neighborhood state, as compared to inland counties in the neighborhood state.

One notable feature of our empirical strategy is that the identification of the border effects ($\beta_3$ and $\beta_4$) does not require any assumption about the exogeneity of law change. As we've discussed in the literature review, enactments of PDMP laws are endogenous. When each state decides to implement ePDMP is a function of many factors, including its regulatory environment, the severity of its opioid crisis, the current political climate, and many others. These factors are highly correlated with the pre-enactment level of sales and mortality and the post-enactment response. If states are more likely to pursue stringent opioid regulations when conditions are bad, $\beta_1$ would underestimate the true impact of the law change. In terms of the estimation of the difference ($\beta_3$) and the spillover effect ($\beta_4$), law changes can be considered as random events.

\section{Results}\label{section:results}
\subsection{PDMP Law and spatial spillover in sales}

The full set of $\beta$ from our main regression is presented in column (5) of Table \ref{tab:main_sales}. We start with a simple two-way fixed effects model in column (1). We replace county fixed effects with the set of state fixed effects in (2) to (5) to estimate the border coefficients. In (2), we replicate the same regression as in (1) to show that changing from county to state fixed effects has no discernible impact on the ePMDP law coefficient. Starting in column (3), we add border status and interact it with ePDMP law to separately estimate the impact of ePDMP law on border counties. To ensure that differences in population characteristics between border and inland county are not driving the identification, in column (4) to (5), we control for county characteristics (average age, \% male, \% separated, education level, mean income, \% low income, and ethnicity). These are variables that the literature has characterized as being influential in driving opioid use and overdose \parencite{wright2014iatrogenic}. In column (5), we add an indicator for whether the nearby state adopted ePDMP for each border county. We repeat the same analysis using the alternative ePDMP enactment dates from PDAPS. The results are documented in in Table \ref{tab:robust_sales} in the Appendix.

\begin{table} [H]
	\centering
	\footnotesize
	\renewcommand\arraystretch{1.1}
	\begin{threeparttable}
		
\caption{Impact of ePDMP laws on opioid sales using Horwitz (2018) modern system operational dates}
\label{tab:main_sales}
	
\begin{tabular} {L{0.25\textwidth}C{0.09\textwidth}C{0.09\textwidth}C{0.09\textwidth}C{0.09\textwidth}C{0.09\textwidth}}
\hline\hline
 &\multicolumn{5}{c}{Dependent variable:}\\\cline{2-6}
 &\multicolumn{5}{c}{Sales per person} \\
 & (1) & (2) & (3) & (4) & (5) \\\hline
 $\beta_1$ - PDMP law & -0.006***  & -0.005***&-0.002&-0.003&-0.003\\
 & (0.001)  & (0.002) & (0.002) & (0.002)& (0.002)\\
 $\beta_2$ - Border county &  & & 0.006*** & 0.004*** & 0.004***\\
 &&& (0.002) & (0.001)& (0.002)\\
 $\beta_3$ - Law x border & && -0.009*** & -0.008***&-0.008***\\
 &&& (0.002) & (0.002)& (0.002)\\
 $\beta_4$ - Nearby law x border&&&&&0.0004\\
 &&&&&(0.002)\\\hline
 County FE & Yes  &&&&\\
 State FE & & Yes & Yes & Yes & Yes \\
 Year FE &  Yes & Yes & Yes & Yes & Yes \\
 Controls &&&& Yes & Yes\\\hline
 Observations &  26,154 & 26,154 & 26,154 & 26,154& 26,154\\
 $R^2$ & 0.000 & 0.458 & 0.458 & 0.519 & 0.519 \\\hline\hline
 
\end{tabular}
	\end{threeparttable}
\end{table}

$\beta_1$ estimates the average effect of the enactment of ePMDP laws on opioid sales. Before adjusting for differential response due to the location of the county, we find that ePDMP reduces opioid sales. The coefficient is consistently negative in all specifications, but only significant before the inclusion of border coefficients. The border coefficient $\beta_2$ is consistently positive from (3) to (5), indicating that border counties start with higher sales as compared to inland counties of the same state. The estimation of $\beta_2$ supports \textit{hypothesis 2a} (spillover framework) over \textit{hypothesis 1a} (state-as-island framework). $\beta_3$, the law and border interaction term, is consistently negative. Although border counties start with higher per person sales, they experience a much larger drop in sales post-ePDMP than inland counties in the same state. The results are consistent with \textit{hypothesis 2b} (spillover framework) that a higher percentage of sales in border counties are diverted elsewhere for non-medical use. Comparing the size of $\beta_1$ across specifications, we see that the estimated impact of ePMDP law on sales is largest in columns (1) and (2) and decreases and loses significance once we interact law with border status. If we don't separately account for abnormal behaviors in the border counties, the coefficients in (1) and (2) over-estimate the effect of the law change on opioid sales in a ``normal'' county. We observe the same pattern using our alternative e-access dates in Table \ref{tab:robust_sales}.

In regression (5) of Table \ref{tab:main_sales}, $\beta_4$ is not well identified. Using our alternative e-access date, $\beta_4$ is significant and positive. We need to be careful in interpreting $\beta_4$ since the coefficient is measured with respect to sales in inland counties of the same state. Suppose A and B are two neighboring states and A experiences a law change. We have tentative evidence that counties in B that border A experience a faster growth (or slower decline) in sales than the inland counties in B. The findings support \textit{hypothesis 2c} (spillover framework) over \textit{hypothesis 1c} (state-as-island framework). Implementation of an ePDMP in one state increases the sales of opioids in border counties of nearby states. 

Putting the coefficients together, border counties start with higher sales, experience a larger decrease if the local state enacts the ePDMP, and an additional increase if the nearby state enacts the ePDMP (only if we use alternative ePDMP dates). When states on both sides of the border adopt ePDMPs, most of the border effects cancel out. As the difference in regulation disappears between states, border counties lose their higher-than-average sales and their significance in cross-border opioid trafficking. In addition, the decrease in sales due to ePDMP laws are driven mostly by decreases in the border counties. The inland counties experience no significant drops in sales once we control for the border-law interactions. In the robustness section, we discuss what impact adoption timing has on how border states react to the enactment of electronic PDMP locally and nearby. 

Translating our coefficients to real terms using Table \ref{tab:main_sales}, if we don't differentiate the border counties from inland counties, (1) shows that the law reduces per person sales by 0.006 MME, which is equivalent to a 5.6\% drop from the national average in 2006. Since only a portion of sales are diverted for non-medical use, a 5.6\% overall decrease is large if we translate it into drops in diversions. when we do account for border status, our estimation shows that the law reduces inland county sales by 0.003 mg in active ingredient on average (2.8\%). In addition, the law reduces the border county's sales by 0.011 mg in active ingredient weight (10.2\%), which is more than three times as much as the drop in inland counties.

\subsection{PDMP Law and spatial spillover in mortality}

We've shown that the adoption of PDMP laws decreases local sales but has spillover effects on nearby states. Ultimately, however, what we care about is the consequences these laws have on actual opioid misuse and overdose. In this section, we use the same econometric specifications to test what impact an ePDMP law enacted in a state have on mortality in local and nearby counties. We expect the mortality results to be less sharp than sales results since there are many intervening factors between access to prescription opioids and opioid overdoses. Spatial spillovers, as identified in the previous section, are one. Substitution toward other alternative drugs is another. The literature has many examples of how restricting access to one drug resulted in substitution toward another potentially more lethal substance (\cite{alpert2018supply}; \cite{zhang2021oxycontin}; \cite{kim2021must}).

\begin{table} [H]
	\centering
	\footnotesize
	\renewcommand\arraystretch{1.1}
	\begin{threeparttable}
		
\caption{Impact of ePDMP laws on opioid mortality using Horwitz (2018) modern system operational date}
\label{tab:main_mortality}
	
\begin{tabular} {L{0.25\textwidth}C{0.09\textwidth}C{0.09\textwidth}C{0.09\textwidth}C{0.09\textwidth}C{0.09\textwidth}}
\hline\hline
 &\multicolumn{5}{c}{Dependent variable:}\\\cline{2-6}
 &\multicolumn{5}{c}{Mortality per 100,000 residents} \\
 & (1) & (2) & (3) & (4) & (5)\\\hline
 $\beta_1$ - PDMP law & -0.217*** & -0.192*** & -0.302*** & -0.318***&-0.279***\\
 &  (0.051) &  (0.069) &  (0.075)  & (0.073)  & (0.074) \\
 $\beta_2$ - Border county &  & & -0.580*** & -0.666*** & -0.763***\\
 &&&  (0.063)  & (0.062) & (0.067)\\
 $\beta_3$ - Law x border & && 0.320*** & 0.366*** & 0.254***\\
 &&& (0.085) & (0.083) & (0.088)\\
 $\beta_4$ - Nearby law x border &&&&& 0.297***\\
 &&&&&(0.076)\\\hline
 County FE & Yes  &&&&\\
 State FE & & Yes & Yes & Yes & Yes\\
 Year FE &  Yes & Yes & Yes & Yes & Yes \\
 Controls &&&& Yes & Yes\\\hline
 Observations &  26,154 & 26,154 & 26,154 & 26,154& 26,154\\
 $R^2$ & 0.000 & 0.283   &  0.285 &  0.318 & 0.318\\\hline\hline
 
\end{tabular}
	\end{threeparttable}
\end{table}

The coefficients on \textit{PDMP law} are straightforward to interpret. Across the specifications, PDMP laws reduce opioid overdose. The reduction is economically significant. Using estimates from column (5), a -0.279 drop per 100,000 people translates into a 12.3\% drop from the national opioid fatality rate in 2006. A negative and significant $\beta_2$ indicates that border counties have a lower level of baseline overdoes rate, which is consistent with our hypothesis that border counties don't abuse as many opioids but export a high percentage of their sales for misuse elsewhere (\textit{hypothesis 2a}). Given that the extra sales originating from border counties are not consumed locally, the adoption of PDMP laws should have no extra impact, if not less, on mortality in these counties. In columns (3) to (5), our estimation of $\beta_3$ is positive and significant. The size of $\beta_3$ is almost as large as $\beta_1$ in all three specifications, suggesting that the adoption of ePDMP has nearly no impact on a border county, which supports \textit{hypothesis 2d}. 

In regression (5), we find $\beta_4$ to be positive and significant, which suggests that the mortality rate in border counties neighboring a state with a new ePDMP law increases faster (or decreases slower) than that in the inland counties of the same state. We get similar findings using the alternative ePDMP dates (Table \ref{tab:robust_mortality}). A positive $\beta_4$ does not support \textit{hypothesis 2e} that nearby enactment of ePDMP has no addition impact on the border counties. While the sales results suggest that people from recently restrictive states cross the state line to acquire opioids from the neighbor county, the mortality results suggest that these people not only shop across state lines, but also stay in the neighbor county to consume these opioids. Validating this mechanism is beyond the scope of the data we have, and we leave it to future researchers.

The differences in the mortality and the sales results are direct evidence that prescription opioids are trafficked across state lines. If opioids sold in each county are consumed locally, the mortality result should mirror that of the sales result. However, we find that border counties start with higher levels of sales but lower levels of mortality. Enactment of ePDMP leads to additional drops in sales in border counties, but fewer drops in mortality. The overall evidence supports the spillover framework over the state-as-island framework. 

\subsection{Effectiveness of alternative PDMP laws}

To check if other PDMP laws have similar spillover effects, we run our main regression using two additional dates: any PMDP dates from \citet{horwitz2018problem} and MA-PDMP dates from \citet{sacks2021can}. The results on sales are documented in Table \ref{tab:rob_dates_sales} in Appendix. The enactments of PDMP and must-access PDMP are not associated with reductions in opioid sales during our sample period. These findings are not conclusive evidence that PMDPs or MA-PDMPs are ineffective in reducing opioid sales. As we've stated in Section 3.2 and shown in Figure \ref{fig:implementation}, our sample period covers very few enactments of PDMPs and MA-PDMPs. Most states had already enacted some version of the PDMP by the start of our sample period, hence we only observe PDMP law change in few states that had been slow in action. Similarly for MA-PDMP, we only observe law change in the few early-mover states. The limited data combined with the endogeneity of adoption means that we do not have enough power to identify the effects of PMDP and MA-PDMP using 2006 to 2014 data.

We identify no border or spillover effect using the two alternative dates. The results suggest that identification of the border and spillover effects is sensitive to using the ``correct'' PDMP law. On the border coefficient, we know from previous regressions that the enactment of ePDMPs on both sides of the border makes the border counties lose their significance in cross-border shopping. Not finding a border effect using PDMP or MA-PDMP dates further validates our main hypotheses. On the spillover effect, if the law itself did not lead to a significant reduction in opioid sales in the first place, there is no reason to expect individuals to cross-border shop. 

\section{Conclusion}\label{section:conclusion}

In this paper, we examined the effects of ePDMP laws on both the states they were enacted in and neighboring states. Following the literature, we find that opioid sales fall in states that adopted electronic access PDMPs. After controlling for border and spillover effects, we estimate that ePDMP laws reduce per-person opioid sales by 5.6\% from the median sales in 2006, a considerable drop because the laws should only affect the fraction of users doctor or pharmacy shopping. We find that the decrease was driven by border counties in particular, where sales decreased 10.2\% post-ePDMP. We also find that ePDMP laws reduce opioid overdoses in a state, with approximately a 12.3\% decrease relative to per-capita mortality in 2006. These findings confirm the understanding in the literature that PDMP laws are effective in curbing the opioid epidemic. 

The decentralized adoption of ePDMPs created opportunities for individuals to cross the state border to acquire opioids from a less restrictive state. Counties on the border are more likely to be destinations for doctor or pharmacy shopping, due to the lower travel cost from other states. Our paper is the first to document a differential pattern in opioid use and a differential response to law changes in counties due to their proximity to the state border. Before the enactment of an ePDMP, border counties have significantly higher opioid sales and lower rates of overdose as compared to inland counties of the same state. When the state adopts an ePDMP, its border counties experience a larger drop in sales and a smaller decrease in mortality. In addition, when the nearby state adopts an ePDMP, we observe a larger increase (or smaller decrease) in sales in counties neighboring the law change state as compared to inland counties in that same state. The spillover effect indicates that the benefits of ePDMPs are partially mitigated because individuals purchase opioids from neighboring states when their state adopts an ePDMP. 

The qualitative differences between border and inland counties in opioid sales and overdose have implications for all studies on the opioid crisis. Previous studies treat each state as an independent market and assume that local opioid sales have a one-to-one mapping to local opioid consumption. This simplifying assumption is the correct one to make in many situations. For example, in the study of the OxyContin reformulation, each state is treated with the same regulatory change. Spillover effects due to preexisting regulatory or cultural differences still exist, but they are irrelevant to measuring the impact of OxyContin reformulation on opioid use. However, in many other situations, where change takes place on a state-by-state basis, treating each state as an independent market may bias the estimation. In the case of PDMP laws, not accounting for cross-border sales overestimates the benefits of the law change. 

The spillovers we have identified in this paper have implications beyond the opioid crisis. We have documented a direct negative externality from having state-based opioid policies instead of a national one. In a counterfactual world where all states adopt electronic access of PDMPs at the same time, all states would get the sales reduction without the increased sales from cross-border trafficking. These findings speak to the advantages and disadvantages of a federalist system. On one hand, decentralization allows each state to experiment and adopt politics based on their own conditions. Information from early adopters could flow to late adopters, thereby providing late adopters with real-world data on policy effectiveness. On the other hand, decentralization kills coordination and there is often a cost in failures to coordinate. Individuals, resources, and businesses are often not confined to one location. Regulatory differences among states allow entities unwilling to comply to move to a different state, thereby offsetting the positive benefit of new regulations.

This study is the second in a series of papers using the ARCOS data to better understand the opioid crisis. The first paper discusses substitution toward generic oxycodone when OxyContin was no longer abusable due to Purdue's reformulation. This paper discusses geographic substitution when obtaining opioids from one location becomes more restrictive due to the enactment of an ePDMP. 

\newpage

\printbibliography

\section{Appendix}

\subsection{Additional Tables}

% \begin{table} [H]
% 	\centering
% 	\footnotesize
% 	\renewcommand\arraystretch{1.1}
% 	\begin{threeparttable}
% 		\input{table_0615/endogeneity}
% 	\end{threeparttable}
% \end{table}

\begin{table} [H]
	\centering
	\footnotesize
	\begin{threeparttable}
		\caption{PDMP Adoption Timing}
\label{tab:date}
	
\begin{tabular} {L{0.2\textwidth}C{0.15\textwidth}C{0.15\textwidth}C{0.15\textwidth}C{0.15\textwidth}}
\hline\hline
State & PDMP & ePDMP (Horwitz) & ePDMP (PDAPS) & MA-PDMP\\\hline
Alabama & 2005-11-01 & 2006-04-01 & 2007-06-28 &  \\
Alaska & 2008-09-01 & 2012-01-01 & 2012-01-01 &  \\
Arizona & 2007-09-01 & 2008-12-01 & 2008-12-01 &  \\
Arkansas & 2013-03-01 & 2013-05-01 & 2013-05-16 &  \\
California & 1939-01-01 & 2009-09-01 & 2009-09-01 &  \\
Colorado & 2005-06-01 & 2008-02-01 & 2008-02-04 &  \\
Connecticut & 2006-10-01 & 2008-07-01 &  & 2015-10-01 \\
Delaware & 2011-09-01 & 2012-08-01 & 2012-08-21 & 2012-03-01 \\
District of Columbia & 2014-02-01 & 2016-10-01 &  &  \\
Florida & 2010-12-01 & 2011-10-01 & 2011-10-17 &  \\
Georgia & 2011-07-01 & 2013-05-01 & 2013-07-01 & 2014-07-01 \\
Hawaii & 1943-01-01 & 2012-02-01 & 1997-01-01 &  \\
Idaho & 1967-01-01 & 2008-04-01 & 1999-06-01 &  \\
Illinois & 1961-01-01 & 2009-12-01 &  &  \\
Indiana & 1997-01-01 & 2007-07-01 & 2004-12-29 & 2014-07-01 \\
Iowa & 2006-05-01 & 2009-03-01 & 2009-03-19 &  \\
Kansas & 2008-07-01 & 2011-04-01 & 2011-04-01 &  \\
Kentucky & 1998-07-01 & 1999-07-01 & 1999-07-01 & 2012-07-01 \\
Louisiana & 2006-07-01 & 2009-01-01 & 2009-01-01 & 2008-01-01 \\
Maine & 2004-01-01 & 2005-01-01 & 2005-01-01 &  \\
Maryland & 2011-10-01 & 2013-12-01 & 2013-12-20 &  \\
Massachusetts & 1992-12-01 & 2011-01-01 & 2011-01-01 & 2014-07-01 \\
Michigan & 1988-01-01 & 2003-01-01 & 2003-01-01 &  \\
Minnesota & 2009-01-01 & 2010-04-01 & 2010-04-15 &  \\
Mississippi & 2006-06-01 & 2008-07-01 & 2005-12-01 &  \\
Missouri &  &  &  &  \\
Montana & 2011-07-01 & 2012-10-01 & 2012-11-01 &  \\
Nebraska & 2011-08-01 & 2017-01-01 & 2011-04-14 &  \\
Nevada & 1996-01-01 & 2011-02-01 & 1997-07-01 & 2007-10-01 \\
New Hampshire & 2012-06-01 & 2014-10-01 & 2014-10-16 & 2016-01-01 \\
New Jersey & 2009-08-01 & 2012-01-01 & 2012-01-05 & 2015-11-01 \\
New Mexico & 2004-07-01 & 2005-08-01 & 2005-08-01 & 2012-09-01 \\
New York & 1972-01-01 & 2013-06-01 &  & 2013-08-01 \\
North Carolina & 2006-01-01 & 2007-07-01 & 2007-10-01 &  \\
North Dakota & 2006-12-01 & 2008-10-01 & 2007-09-01 &  \\
Ohio & 2005-05-01 & 2006-10-01 & 2006-10-02 & 2012-03-01 \\
Oklahoma & 1991-01-01 & 2006-07-01 & 2006-07-01 & 2011-03-01 \\
Oregon & 2009-07-01 & 2011-09-01 & 2011-09-01 &  \\
Pennsylvania & 1972-01-01 & 2016-08-01 &  &  \\
Rhode Island & 1978-01-01 & 2012-09-01 & 2012-07-01 & 2016-06-01 \\
South Carolina & 2006-06-01 & 2008-02-01 & 2008-09-01 &  \\
South Dakota & 2010-03-01 & 2012-03-01 & 2012-03-01 &  \\
Tennessee & 2003-01-01 & 2010-01-01 & 2007-01-01 & 2013-07-01 \\
Texas & 1981-09-01 & 2012-08-01 & 2012-06-30 &  \\
Utah & 1995-07-01 & 2006-01-01 & 1997-01-01 &  \\
Vermont & 2008-06-01 & 2009-01-01 & 2009-04-01 & 2015-05-01 \\
Virginia & 2003-09-01 & 2006-01-01 & 2006-06-01 & 2015-07-01 \\
Washington & 2011-08-01 & 2012-01-01 & 2012-01-04 &  \\
West Virginia & 1995-06-01 & 2013-05-01 & 2004-12-01 & 2012-06-01 \\
Wisconsin & 2010-06-01 & 2013-06-01 & 2013-06-01 &  \\
Wyoming & 2003-07-01 & 2013-07-01 & 2004-10-01 &\\ \hline\hline
 
\end{tabular}

\begin{tablenotes}
	\small
	\item \textit{Notes:} Date in the first column is the enactment/legislated start date for any PDMP from \citet{horwitz2018problem}. Date in the second column is the modern system operational date from \citet{horwitz2018problem}. Date in the third column is the electronic access dates from PDAPS. Date in the forth column is the must-access PDMP date from \citet{sacks2021can}.
\end{tablenotes}
	\end{threeparttable}
\end{table}

\begin{table} [H]
	\centering
	\footnotesize
	\renewcommand\arraystretch{1.1}
	\begin{threeparttable}
		
\caption{Impact of ePDMP laws on opioid sales using PDAPS dates}
\label{tab:robust_sales}
	
\begin{tabular} {L{0.25\textwidth}C{0.09\textwidth}C{0.09\textwidth}C{0.09\textwidth}C{0.09\textwidth}C{0.09\textwidth}}
\hline\hline
 &\multicolumn{5}{c}{Dependent variable:}\\\cline{2-6}
 &\multicolumn{5}{c}{Sales per person} \\
 & (1) & (2) & (3) & (4) & (5) \\\hline
 $\beta_1$ - PDMP law & -0.010***  & -0.011***&-0.008**&-0.008**&-0.008**\\
 & (0.001)  & (0.002) & (0.002) & (0.002)& (0.002)\\
 $\beta_2$ - Border county &  & & 0.006*** & 0.005***&0.003*\\
 &&& (0.002) & (0.001)& (0.002)\\
 $\beta_3$ - Law x border & && -0.009*** & -0.008***&-0.009***\\
 &&& (0.002) & (0.002)& (0.003)\\
 $\beta_4$ - Nearby law x border&&&&& 0.003*\\
 &&&&&(0.002)\\\hline
 County FE & Yes &&&&\\
 State FE & & Yes & Yes & Yes & Yes \\
 Year FE &  Yes & Yes & Yes & Yes & Yes \\
 Controls &&&& Yes & Yes\\\hline
 Observations &  26,154 & 26,154 & 26,154 & 26,154& 26,154\\
 $R^2$ & 0.005 & 0.459 & 0.459 & 0.520 & 0.520\\\hline\hline
 
\end{tabular}

\begin{tablenotes}
	\small
	\item \textit{Notes:} We run the same regressions as Table \ref{tab:main_sales} using alternative ePMDP dates. The results are very similar to our main findings: ePDMP reduces sales, but the reduction is less once we control for border counties; border counties have higher level of sales and they experience sharper decline when ePDMP is enacted; enactment of ePDMP in nearby states increases sales in border counties of the local state.
\end{tablenotes}
	\end{threeparttable}
\end{table}

\begin{table} [H]
	\centering
	\footnotesize
	\renewcommand\arraystretch{1.1}
	\begin{threeparttable}
		
\caption{Impact of ePDMP laws on opioid mortality using PDAPS dates}
\label{tab:robust_mortality}
	
\begin{tabular} {L{0.25\textwidth}C{0.09\textwidth}C{0.09\textwidth}C{0.09\textwidth}C{0.09\textwidth}C{0.09\textwidth}}
\hline\hline
 &\multicolumn{5}{c}{Dependent variable:}\\\cline{2-6}
 &\multicolumn{5}{c}{Mortality per 100,000 residents} \\
 & (1) & (2) & (3) & (4) & (5)\\\hline
 $\beta_1$ - PDMP law & -0.419*** & -0.391*** & -0.432*** & -0.457***&-0.431***\\
 &  (0.052) &  (0.071) &  (0.075)  & (0.073)  & (0.074) \\
 $\beta_2$ - Border county &  & & -0.488*** & -0.594*** & -0.682***\\
 &&&  (0.062)  & (0.062) & (0.070)\\
 $\beta_3$ - Law x border & && 0.146* & 0.234*** & 0.173*\\
 &&& (0.088) & (0.086) & (0.089)\\
 $\beta_4$ - Nearby law x border &&&&& 0.210***\\
 &&&&&(0.075)\\\hline
 County FE & Yes  &&&&\\
 State FE & & Yes & Yes & Yes & Yes\\
 Year FE &  Yes & Yes & Yes & Yes & Yes \\
 Controls &&&& Yes & Yes\\\hline
 Observations &  26,154 & 26,154 & 26,154 & 26,154& 26,154\\
 $R^2$ & 0.001 & 0.283   &  0.286&  0.318 & 0.319\\\hline\hline
 
\end{tabular}

\begin{tablenotes}
	\small
	\item \textit{Notes:} We run the same regressions as Table \ref{tab:main_mortality} using alternative ePDMP dates. Again, the results are almost identical to our main findings. Enactment of ePDMP laws reduces overdose. Border counties start with lower opioid mortality but experience almost no drop when the state enacts ePDMP. Nearby enactment of ePDMP has a spillover effect on the mortality in the border counties of the local state.
\end{tablenotes}
	\end{threeparttable}
\end{table}

\begin{table} [H]
	\centering
	\footnotesize
	\renewcommand\arraystretch{1.1}
	\begin{threeparttable}
		
\caption{Impact of ePDMP laws on opioid sales using any PDMP dates, e-access dates, and must-access dates}
\label{tab:rob_dates_sales}
	
\begin{tabular} {L{0.02\textwidth}L{0.25\textwidth}C{0.09\textwidth}C{0.09\textwidth}C{0.09\textwidth}C{0.09\textwidth}C{0.09\textwidth}}
\hline\hline
 &&\multicolumn{5}{c}{Dependent variable:}\\\cline{3-7}
 &&\multicolumn{5}{c}{Sales per person} \\
 && (1) & (2) & (3) & (4) & (5) \\\hline
 
 \multicolumn{7}{l}{\textit{(A) Any PDMP}} \\
 &$\beta_1$ - PDMP law & 0.006***  & 0.005*** & 0.005** & 0.004 & 0.003\\
 && (0.001)  & (0.002) & (0.003) & (0.002)& (0.002)\\
 &$\beta_2$ - Border county &  & & 0.002 & 0.003 & 0.002\\
 &&&& (0.003) & (0.003)& (0.003)\\
 &$\beta_3$ - Law x border & && 0.0001 & -0.004&-0.004\\
 &&&& (0.003) & (0.003)& (0.003)\\
 &$\beta_4$ - Nearby law x border&&&&&-0.001\\
 &&&&&&(0.002)\\

 \multicolumn{7}{l}{\textit{(B) Electronic access PDMP (main regression)}} \\
 &$\beta_1$ - PDMP law & -0.006***  & -0.005***&-0.002&-0.003&-0.003\\
 && (0.001)  & (0.002) & (0.002) & (0.002)& (0.002)\\
 &$\beta_2$ - Border county &  & & 0.006*** & 0.004*** & 0.004***\\
 &&&& (0.002) & (0.001)& (0.002)\\
 &$\beta_3$ - Law x border & && -0.009*** & -0.008***&-0.008***\\
 &&&& (0.002) & (0.002)& (0.002)\\
 &$\beta_4$ - Nearby law x border&&&&&0.0004\\
 &&&&&&(0.002)\\

 \multicolumn{7}{l}{\textit{(C) Must access PDMP}} \\
 &$\beta_1$ - PDMP law & 0.004*** & 0.003 & -0.001& 0.001 & 0.001\\
 && (0.001)  & (0.003) & (0.003) & (0.003)& (0.003)\\
 &$\beta_2$ - Border county &  & & 0.001 & 0.0001 & -0.001\\
 &&&& (0.001) & (0.001)& (0.001)\\
 &$\beta_3$ - Law x border & && 0.012** &0.006 &0.006 \\
 &&&& (0.005) & (0.004)& (0.004)\\
 &$\beta_4$ - Nearby law x border&&&&&0.010***\\
 &&&&&&(0.003)\\

 \hline
 &County FE & Yes  &&&&\\
 &State FE & & Yes & Yes & Yes & Yes \\
 &Year FE &  Yes & Yes & Yes & Yes & Yes \\
 &Controls &&&& Yes & Yes\\
 &Observations &  26,154 & 26,154 & 26,154 & 26,154& 26,154\\\hline\hline
 
\end{tabular}
	\end{threeparttable}
\end{table}

\newpage
\subsection{Additional Figures}
	
	\begin{figure} [H]
		\centering
	
		\begin{minipage}{0.9\textwidth}\
        \includegraphics[width=\linewidth]{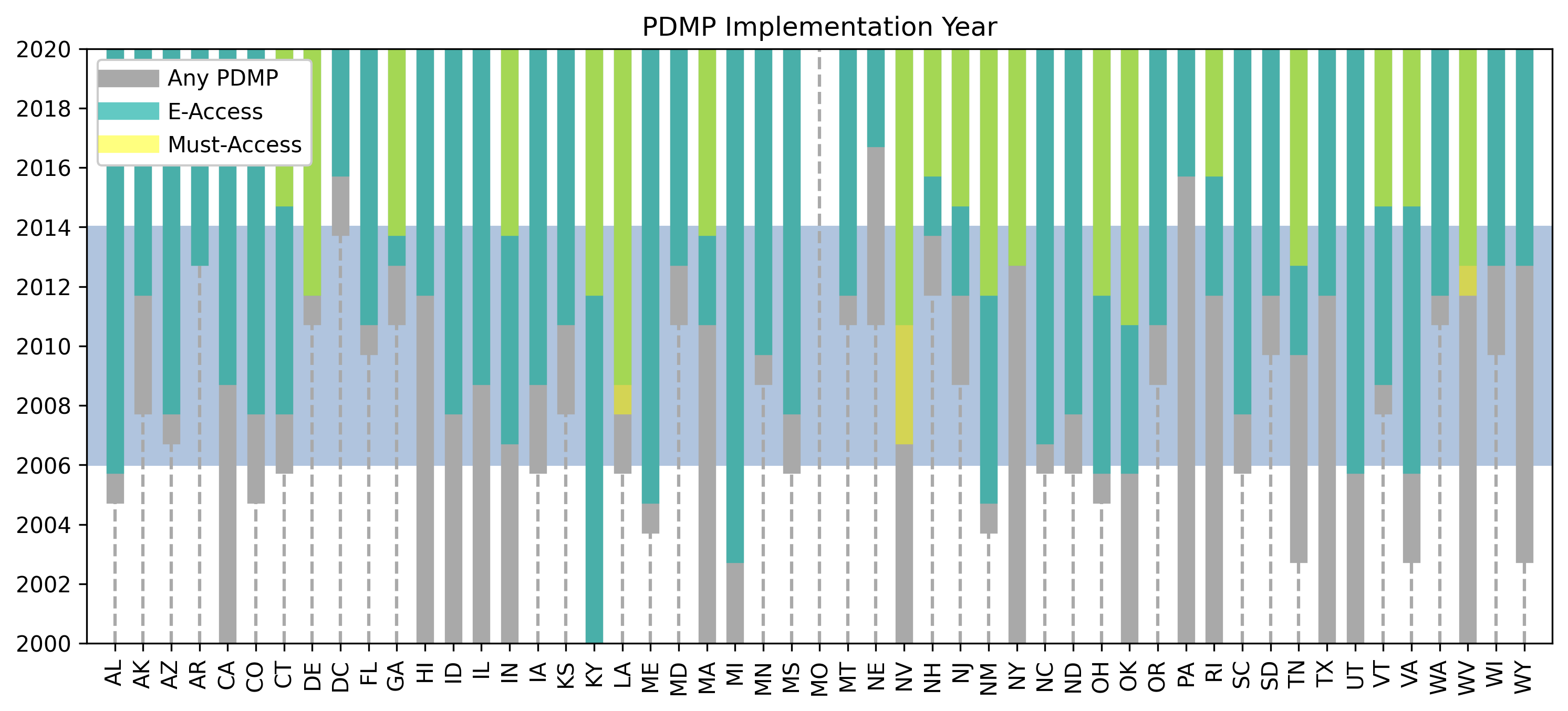}
        {\footnotesize \textit{Note:} The horizontal blue rectangle marks our sample period (2006 to 2014). For ePDMP, 9 states adopted before the start of our sample period, 16 states adopted in the first half of our sample, 18 states adopted in the second half, and 8 states had not adopted by the end of our sample period. \par}
        \end{minipage}
		
		\caption{PDMP implementation dates by state}
		\label{fig:implementation}
	\end{figure}
	
	\begin{figure} [H]
		\centering
	
		\begin{minipage}{0.9\textwidth}\
        \includegraphics[width=\linewidth]{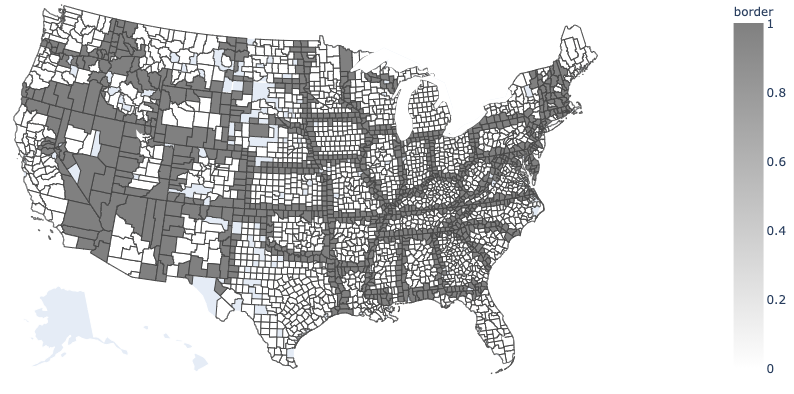}
        \end{minipage}
		
		\caption{Map of border vs inland counties}
		\label{fig:map_border}
    \end{figure}

    \begin{figure} [H]
		\centering
	
		\begin{minipage}{0.8\textwidth}\
        \includegraphics[width=\linewidth]{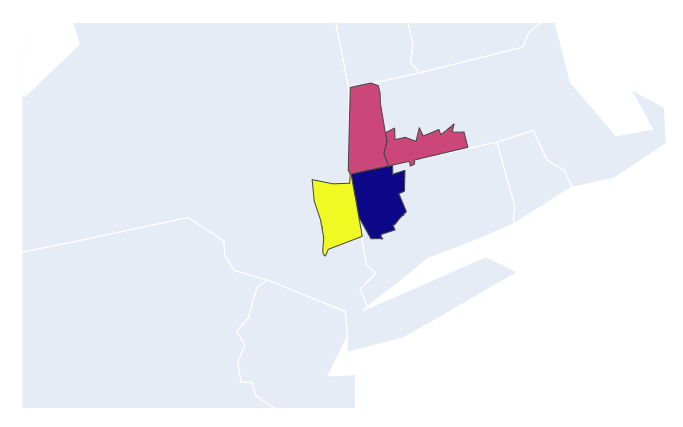}
        {\footnotesize \textit{Note:} This picture illustrates how we calculate nearby law for Litchfield County, Connecticut (blue) in 2012 . The Litchfield County borders three counties from nearby states: Dutchess County from New York (yellow), and Birkshire County and Hampden County from Massachusetts (pink). In 2012, the state of New York has not adopted ePDMP and the state of Massachusetts has. To calculate nearby law for Litchfield County, we sum up the population nearby with no ePDMP ($294,000$) and the population nearby with ePDMPs ($125,000 + 466,000 = 591,000$). Since more people nearby live in counties with ePDMP, nearby law for Litchfield County in 2012 is 1. \par}
        \end{minipage}
		
		\caption{Calculating nearby ePMDP status for counties bordering several states}
		\label{fig:nearby_example}
    \end{figure}

    \begin{figure} [H]
		\centering
	
		\begin{minipage}{0.6\textwidth}\
        \includegraphics[width=\linewidth]{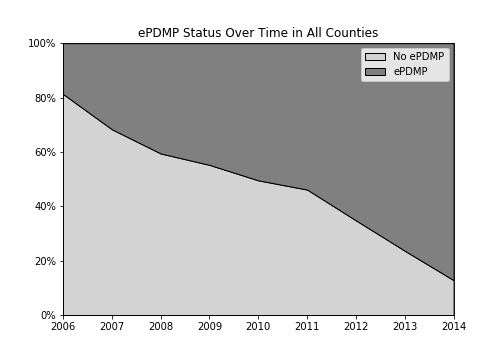}
        \end{minipage}
		
		\caption{ePDMP adoption over time in all counties}
		\label{fig:pdmp_overtime}
    \end{figure}

    \begin{figure} [H]
		\centering
	
		\begin{minipage}{0.6\textwidth}\
        \includegraphics[width=\linewidth]{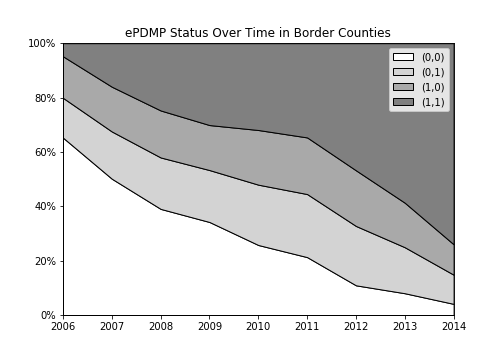} {\footnotesize \textit{Note:} For the ease of reference, we use the (my law, nearby law) syntax to denote the ePDMP status of a border county. A border county of $(0,0)$ has no ePDMP law and no nearby ePDMP law and its cross-state neighbors also do not have one; a border county of $(1,0)$ has an ePDMP law himself but not nearby; a border county of $(0,1)$ does not have a law himself it's nearby state does; and a border county of $(1,1)$ has an ePDMP law itself and so do its out-of-state neighbors. \par}
        \end{minipage}
		
		\caption{ePDMP adoption over time in border counties}
		\label{fig:pdmp_border_overtime}
    \end{figure}

    \begin{figure} [H]
		\centering
	
		\begin{minipage}{0.8\textwidth}\
        \includegraphics[width=\linewidth]{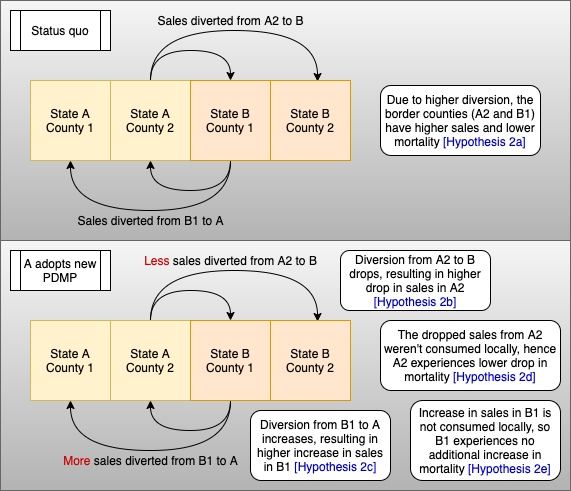}
        \end{minipage}
		
		\caption{Visual presentation of the spillover framework}
		\label{fig:hyp}
    \end{figure}

\end{document}